\documentclass[preprint,nofootinbib]{revtex4-1}
 \usepackage{graphicx,psfrag,epsfig}
\newcommand{\beq}{\begin{equation}}
\newcommand{\eeq}{\end{equation}}

\newcommand{\bea}{\begin{eqnarray}}
\newcommand{\eea}{\end{eqnarray}}

\newcommand{\ea}{{\it et al.}}

\begin{document}
\begin{flushright}
UMD-PP-10-016\\
IFT-UAM/CSIC-10-53\\
FTUAM-10-14\\
October, 2010\
\end{flushright}
\vspace{0.3in}
\title {\LARGE Leptogenesis with TeV Scale Inverse Seesaw in $SO(10)$}
\author {\bf Steve Blanchet$^{a,b}$, P. S. Bhupal Dev$^a$ and
R. N. Mohapatra$^a$\\}
\affiliation{$^a$ Maryland
Center for Fundamental Physics and Department of Physics,
University of Maryland, College Park, MD 20742, USA\\
$^b$ Instituto de F\'isica
Te\'orica, IFT-UAM/CSIC, Nicolas Cabrera 15, UAM, Cantoblanco,
28049 Madrid, Spain}
\vspace{0.5in}
\begin{abstract}
We discuss leptogenesis within a TeV-scale inverse seesaw model for
neutrino masses where the seesaw structure is guaranteed by an
$SO(10)$ symmetry. Contrary to the TeV-scale type-I gauged seesaw,
the constraints imposed by successful leptogenesis in these models
are rather weak and allow for the extra gauge bosons $W_R$ and $Z'$
to be in the LHC accessible range. The key differences in the
inverse seesaw compared to the type I case are: (i) decay and
inverse decay rates larger than the scatterings involving extra
gauge bosons due to the large Yukawa couplings and (ii) the
suppression of the washout due to very small lepton number
breaking.
\end{abstract}
\maketitle

\section{Introduction}

One of the attractive features of the seesaw mechanism for neutrino
masses~\cite{seesaw} is that it provides a way to understand the
origin of matter in the Universe via leptogenesis~\cite{fy} (for a
recent review, see Ref.~\cite{review}). In the vanilla framework of
leptogenesis where right-handed (RH) neutrino masses are
hierarchical, it is well known that the lightest RH neutrino needs
to be rather heavy, around $10^{9}$~GeV or
higher~\cite{Davidson:2002qv}. These scales are however beyond the
reach of collider experiments, e.g. the CERN Large Hadron Collider
(LHC). On the other hand, from the point of view of the seesaw model
itself, one can envisage the new physics scale to be anywhere
between TeV to $10^{14}$~GeV. It is well known that the scale of
leptogenesis can be lowered to the TeV scale if one allows the RH
neutrinos to be quasi-degenerate~\cite{Pilaftsis}. However, first
the quasi-degeneracy should be motivated, and second if the scale of
the RH neutrinos is to be explained by the breaking of some gauge
symmetry, what is the impact of the latter on leptogenesis?

Two classes of seesaw models are of interest in this connection: the
usual type-I seesaw~\cite{seesaw}, and the inverse seesaw~\cite{mv}.
In both classes of models, a higher gauge symmetry, e.g. $B-L$, is
usually called for to make the model ``natural''. In addition to
providing a compelling reason for the inclusion of the RH neutrinos
to guarantee anomaly cancellation, in the type-I case it can be used
to understand why the seesaw scale is so much lower than the Planck
scale, whereas, in the inverse seesaw case, it stabilizes the zeros
in the $(\nu, N, S)$ mass matrix that leads to the doubly-suppressed
seesaw formula.

An attractive gauge symmetry that embeds the $B-L$ symmetry and also
provides a way to understand the origin of parity violation in
low-energy weak interactions is the Left-Right (LR) gauge group
$SU(2)_L\times SU(2)_R\times U(1)_{B-L}$~\cite{LR}. An important
question that arises in these models is: What is the scale of parity
invariance? In particular, if it is in the TeV range and if at the
same time leptogenesis generates the desired matter-anti-matter
asymmetry, then the LHC could be probing neutrino mass physics as
well as shed light on one of the deepest mysteries of cosmology.

Since Sakharov's out-of-equilibrium condition~\cite{sakharov}
must be satisfied in
order to generate a baryon asymmetry, the existence of new
interactions inherent to the LR models make it a nontrivial task to
check whether a TeV-scale $W_R$ is indeed compatible with
leptogenesis as an explanation of the origin of matter.
Specifically, the efficiency of leptogenesis crucially depends on
the number of RH neutrinos that decay out of equilibrium to
produce a leptonic asymmetry. This number is set by two things:
First, it depends on the relative magnitudes of the decay rate and
the ($C\!P$-conserving) gauge scattering rates of the RH neutrino,
since this can lead to a dilution of the number of ``useful'' RH
neutrinos. Second, the washout processes, primarily inverse
decays, should drop out of equilibrium early enough, otherwise the
number of RH neutrinos gets suppressed at an exponential rate.

These issues have been analyzed for the type-I case within LR
symmetric models~\cite{Frere:2008ct} as well as $B-L$
models~\cite{blanchet}. It was found that for the full LR models
with TeV-scale parity restoration and RH neutrino masses, gauge
scattering rates induced by $W_R$ exchange largely dominate the
decay and inverse decay rates because the Yukawa couplings are
small for the standard type-I seesaw at the TeV scale. These facts
lead to a huge dilution of the number of RH neutrinos which decay
out of equilibrium and in a $C\!P$ asymmetric manner. Moreover,
the gauge scattering interactions also wash out lepton number at a
very large rate, much larger than the inverse decays. Altogether,
these two effects lead to a very stringent constraint on the mass
scale of $W_R$ for successful leptogenesis, $M_{W_R}\geq
18$~TeV~\cite{Frere:2008ct}, which would imply that the discovery
of a $W_R$ at the LHC is incompatible with thermal leptogenesis as
the origin of matter. On the other hand, in the case of a simple
$B-L$ theory, successful leptogenesis only implies that
$M_{Z'}\geq 2.5$~TeV in the ``collider-friendly'' region of
parameter space where the RH neutrino mass is less than half the
$Z'$ mass~\cite{blanchet}\footnote{For a discussion of low scale
leptogenesis in an $SO(10)$ model where only the doubly charged
Higgs boson is in the TeV range, see Ref.~\cite{majee}. }. We note
that there exist bounds on the $W_R$ mass from low energy
observations~\cite{bounds} and they allow $W_R$ mass to be as low
as 2.5 TeV.

In this paper, we have analyzed the leptogenesis constraints on
the recently proposed TeV-scale LR model within a unified
supersymmetric $SO(10)$ framework~\cite{dev} where neutrino masses
arise from an inverse seesaw mechanism\footnote{For other low-scale 
leptogenesis scenarios in inverse-seesaw-related frameworks, 
see Ref.~\cite{concha}.}. Two features distinguish
the inverse seesaw mechanism from the type-I seesaw: (i) the Dirac
Yukawa couplings of the RH neutrino $N$ can be much larger ($\sim
10^{-1}-10^{-2}$) than for the type-I case (where they are
typically of order $\sim 10^{-6}$ for TeV-scale RH neutrino
masses) and (ii) the lepton-number-violating parameter (the
Majorana mass $\mu$ of the left-right singlet lepton $S$, which
measures the ``pseudo-Diracness'' of $N$) is much smaller than the
Dirac mass of $N$. As a result, first, the decay rate of $N$ can
be much larger than the $W_R$ exchange scattering rate at the
baryogenesis epoch, and second, the wash-out processes are
suppressed by the small Majorana mass $\mu$. Consequently, we find
that both the $W_R$ and $Z'$ can be in the TeV range and hence
accessible at the LHC. This is the main result of our paper, and
it should make the case for searching the $W_R$ and $Z'$ at LHC
stronger~\cite{LHCWR}.

This paper is organized as follows: in Section II, we summarize the
LR inverse seesaw model and give the Dirac Yukawa couplings as well
as the various lepton-number-violating parameters as constrained by
$SO(10)$ unification~\cite{dev}. In Section III, we present a
generic discussion of leptogenesis in this class of inverse seesaw
models; in Section IV, we present the numerical results for our
model. Finally, we summarize our findings and conclude in
Section V. In Appendix A, we present a new scenario for gauge coupling
unification (different from that discussed in Ref.~\cite{dev}) in these
models where the relative magnitudes of $W_R$ and $Z'$ masses can be
unrelated. In Appendix B, we give the analytical expressions of
the $C\!P$-asymmetry in the inverse seesaw model for some special
cases.
\section{Left-Right Inverse Seesaw Parameters in $SO(10)$}
The implementation of the inverse seesaw mechanism~\cite{mv}
requires, in addition to the usual Standard Model (SM) singlet RH
neutrinos $N_i$ ($i=1,2,3$ for three generations) as in the
typical type-I seesaw, three extra SM gauge singlet fermions $S_i$
coupled to the RH neutrinos through the lepton-number-conserving
couplings of the type ${N}S$, while the traditional RH neutrino
Majorana mass term is forbidden by the lepton number
symmetry\footnote{If we include higher dimensional terms in the
theory, they can induce an $NN$ Majorana mass term but its
magnitude is of order $v^2_{BL}/M_{\rm Pl}\sim 10^{-13}$ GeV
and is too small to affect our discussion.}. In the low energy
theory, dominant lepton number breaking arises only from the
self-coupling term $SS$. The neutrino mass Lagrangian in the
flavor basis is given by
\begin{eqnarray}
    {\cal L}_{\rm mass} = {\nu}^TC^{-1}M_DN+{N}^TC^{-1}M_NS+
    \frac{1}{2}S^TC^{-1}\mu S+~{\rm h.c.},
\end{eqnarray}
where $\mu$ is a complex symmetric $3\times 3$ mass matrix
containing all the lepton-number-violating parameters, and $M_D$ and
$M_N$ are $3\times 3$ mass matrices representing the Dirac
mass terms in the $\nu$--$N$ and $N$--$S$ sectors, respectively. In
the basis $\{\nu,N,S\}$, the full $9\times 9$ neutrino mass matrix
is then given by
\begin{eqnarray}
    {\cal M}_\nu = \left(\begin{array}{ccc}
        0 & M_D & 0\\
        M_D^T & 0 & M_N\\
        0 & M_N^T & \mu
    \end{array}\right)\, .
    \label{eq:bigmass}
\end{eqnarray}
The lepton-number-violating entries in the $\mu$ matrix have to be
much smaller than the Dirac neutrino masses in order to fit the
light neutrino masses, as observed in neutrino oscillation
experiments. In fact, the light neutrino mass matrix can be cast
in a seesaw-like form in the limit $\mu \ll M_D \ll M_N$:
\begin{eqnarray}
    m_\nu \simeq M_DM_N^{-1}\mu\left(M_N^T\right)^{-1}M_D^T \equiv
    F\mu F^T,
    \label{eq:lhmass}
\end{eqnarray}
to leading order in $F=M_DM_N^{-1}$.  As expected, in the limit
$\mu\to 0$, which corresponds to unbroken lepton number, we
recover the massless neutrinos of the SM. We note that this
smallness of the $\mu$-parameter peculiar to the inverse seesaw
models allows for a neutrino mass fit even with TeV-scale RH
neutrino mass and large Dirac mass terms. Theoretically, smallness
of the $\mu$-term could be explained in extra dimensional brane
world models if the lepton number is broken in a separate brane
from the standard model brane~\cite{arkani-hamed}.

As shown in Ref.~\cite{dev}, in order to embed a TeV-scale inverse
seesaw mechanism into a generic $SO(10)$ model, we need to break the
$B-L$ gauge symmetry by {\bf 16}-Higgs fields at the TeV scale,
whereas the $SO(10)$ symmetry is broken down to the LR symmetric
gauge group $SU(3)_c\times SU(2)_L\times SU(2)_R\times U(1)_{B-L}$
at the GUT-scale by {\bf 45} and {\bf 54}-Higgs fields; finally, the
SM symmetry is broken at the weak scale by {\bf 10}-Higgs fields. As
in the usual $SO(10)$ models, the three generations of quark and
lepton fields are assigned to three {\bf 16}-dimensional spinor
representations, and correspondingly, we add three $SO(10)$ singlet
matter fields ${\bf 1}_i$  (they can be identified with the $S_i$ fields above)
to implement the inverse seesaw mechanism.

As discussed in Ref.~\cite{dev}, we need at least two ${\bf 10}_H$
fields to have a realistic fermion mass spectrum; we also need two
${\bf 45}_H$ fields, one for symmetry breaking at the GUT scale and
another to give rise to the vectorlike color triplets at the
TeV-scale as required by coupling unification constraints.
Similarly, we need only the $SU(2)_R$ doublet fields of ${\bf 16}_H$
and no $SU(2)_L$ fields for unification\footnote{An alternative choice of
Higgs fields which also consistently leads to coupling unification in this
scenario is presented in Appendix A.}. With this minimal set of
Higgs fields, the most general Yukawa superpotential is given by
\begin{eqnarray}
    W_Y &=& h_{aij}{\bf 16}_i{\bf 16}_j{\bf 10}_{H_a}
    +\frac{f_{aij}}{M^2}{\bf 16}_i{\bf 16}_j{\bf 10}_{H_a}
    {\bf 45}_H{\bf 45}'_H
    +\frac{f'_{aij}}{M}{\bf 16}_i{\bf 16}_j{\bf 10}_{H_a}{\bf 45}_H
    \nonumber\\
    &&
    +f''_{ijk}{\bf 16}_i{\bf 1}_j\overline{\bf 16}_{H_k}+\mu_{ij}{\bf 1}_i
    {\bf 1}_j,
\end{eqnarray}
where the first term is the usual Yukawa coupling term, the second
and third terms are higher-dimensional terms, and the last two terms
give rise to the inverse seesaw mechanism. As already pointed out in
Ref.~\cite{dev}, it is sufficient to keep only one of the
higher-dimensional operators, usually the ${\bf 16}\cdot {\bf
16}\cdot{\bf 10}\cdot{\bf 45}\cdot{\bf 45'}$ term, whose fully
antisymmetric combination acts as an effective ${\bf 126}_H$
operator, in order to obtain a realistic fermion mass spectrum at
the GUT scale, and hence for simplicity, we will assume all the $f'$-couplings
to be zero; keeping this term does not affect our discussion below
\footnote{The $f'$ term has two effective contributions -- one of ${\bf 10}$
-Higgs
type and another of ${\bf 120}$-Higgs type. The effective ${\bf 10}$ coupling can be
absorbed into the first term, and since the ${\bf 120}$ coupling is
antisymmetric in generation indices,
it only contributes to the off-diagonal elements in fermion mass matrices.
Hence, a non-zero $f'$ coupling could only slightly modify the specific
structure of the Dirac neutrino mass matrix, without
changing any of the main results of the paper.}.

The $B-L$ symmetry is broken when the ${\overline{\bf 16}}_H$-field
acquires a vacuum expectation value (VEV) and the $N$--$S$ sector RH
neutrino mass matrix is given by
\begin{eqnarray}
    M_{N_{ij}} = v_Rf''_{ij},
\end{eqnarray}
where $v_R$ is the VEV of ${\overline{\bf 16}}_H$ and is of order
TeV for the low-scale $B-L$ breaking models considered here. All the
other fermion masses are generated when the SM symmetry is broken at
the weak scale by the ${\bf 10}_H$ VEVs. We consider here only the
model (A) of Ref.~\cite{dev} where the VEV patterns of the two ${\bf
10}_H$ fields are given by
\begin{eqnarray}
    \langle \Phi_1\rangle = \left(\begin{array}{cc}
        v_d & 0\\
        0 & 0
    \end{array}\right),~ ~ ~ ~
    \langle \Phi_2\rangle = \left(\begin{array}{cc}
        0 & 0\\
        0 & v_u
    \end{array}\right),
\end{eqnarray}
and the fermion mass matrices are given by
\begin{eqnarray}
    && M_u = \tilde{h}_u+\tilde{f}, ~ ~ ~ ~
       M_d = \tilde{h}_d+\tilde{f},\nonumber\\
       && M_e = \tilde{h}_d-3\tilde{f}, ~ ~ ~ ~
       M_D = \tilde{h}_u-3\tilde{f},
       \label{eq:massfit}
\end{eqnarray}
where in the notation of Ref.~\cite{dev}, $\tilde{h}_{u,d}\equiv
v_{u,d}h_{u,d}$ and $\tilde{f}\equiv v_uf_u = v_df_d$. Using the
renormalization group evolution of the fermion masses in the LR
model, we obtain the GUT-scale fermion masses starting from the
experimentally known weak scale values, and using these mass
eigenvalues, we obtain a fit for the Yukawa coupling matrices at
the GUT scale, from which we can get the structure of the Dirac
neutrino mass matrix.~ 
 Here, as an example,
we quote the result for $\tan\beta\equiv v_u/v_d=10$~\cite{dev}:
\begin{eqnarray}
    M_D = \left(\begin{array}{ccc}
        0.0111 & 0.0384-0.0103{\rm i} & 0.038-0.4433{\rm i}\\
        0.0384+0.0103{\rm i} & 0.2928 & 1.8623+0.0002{\rm i}\\
        0.038+0.4433{\rm i} & 1.8623-0.0002{\rm i} & 77.7573
    \end{array}\right)~{\rm GeV}.
\end{eqnarray}
With this Dirac neutrino mass, we can easily fit the observed neutrino
oscillation data by fixing the singlet mass matrix $\mu$ in
Eq.~(\ref{eq:lhmass}). As an example, for a normal hierarchy of neutrino
masses, and assuming a diagonal structure for the RH neutrino mass matrix
$M_N$ with eigenvalues (3.5, 3, 1) TeV, we can fit the observed
$2\sigma$ neutrino oscillation data~\cite{fogli} for the following choice of
$\mu$:
\begin{eqnarray}
    \mu = \left(\begin{array}{ccc}
        -17.4226+0.3098{\rm i} & 2.1033-0.0590{\rm i} & -0.0136+0.0288{\rm i} \\
        2.1033-0.0590{\rm i} & -0.2585+0.0097{\rm i} & 0.0016-0.0035{\rm i}\\
        -0.0136+0.0288{\rm i} & 0.0016-0.0035{\rm i} & 3.6\times 10^{-5}+
        4.6\times 10^{-5}{\rm i}
\end{array}\right)~{\rm GeV}.
\label{eq:murhs}
\end{eqnarray}
\section{Leptogenesis in Left-Right Inverse Seesaw Models}

In this section we summarize the main features of leptogenesis
within the class of LR inverse seesaw models discussed above. We
also wish to note that while we have used the $SO(10)$ framework
to make the results definite and somewhat more predictive, our
discussion applies also to the case with TeV-scale Left-Right
symmetry without grand unification. In what follows, we will
partially follow the discussion presented in
Ref.~\cite{Blanchet:2009kk}.

As discussed in the introduction, a crucial difference of the
inverse seesaw from the usual seesaw is the dependence on a new
mass matrix, $\mu$, which can lead to the result that no
matter what the ratio of the mass scales $M_D/M_N$ is, the
lightness of the left-handed neutrinos can always be explained by
small $\mu$ entries. In other words, the inverse seesaw makes it
possible to have at the same time large Dirac masses and low, say
TeV-scale, RH neutrino masses, and it still can explain why
neutrinos are light. This is directly connected to the fact that,
in the limit $\mu\to 0$, lepton number is conserved, and therefore
neutrino masses vanish, as in the SM. This is a crucial difference
from the case of TeV scale type I seesaw.

An interesting question to ask is  how leptogenesis is affected
by this distinctive feature of inverse seesaw models. We expect that
the lepton-number-violating washout will go to zero in the limit of
vanishing $\mu$. As a matter of fact, as explicitly shown in
Ref.~\cite{Blanchet:2009kk}, the all-important $\Delta L=2$ washout
process $\ell \Phi \to \bar{\ell}\Phi^{\dagger}$ vanishes as
$\delta_i^2$, with
\begin{eqnarray}
\delta_i=\frac{|M_i-M_j|}{\Gamma_i} \simeq
\frac{\mu_{ii}}{\Gamma_i}\,
\label{eq:deltai},
\end{eqnarray}
where  $\Gamma_i$ is the total decay rate of $N_i$ into lepton and
Higgs (and antiparticles), and $M_{i,j}$ are the masses of the
quasi-Dirac RH neutrino pair $N_{i,j}$ (with $\Gamma_i\simeq
    \Gamma_j$). Note that we denote by $M_i~ (i=1,\ldots,6)$ the heavy neutrino
    mass eigenvalues. As shown in the Appendix, the leading order
    contribution to the mass splitting for each quasi-Dirac pair comes
    from the diagonal elements of the $\mu$ matrix. Therefore, as
    expected, the washout tends to zero in the limit of vanishing $\mu$.
    The suppression of the washout can be shown to occur through the
    destructive interference of one member of a quasi-Dirac pair with
    the other~\cite{Blanchet:2009kk}. It is instructive to show
    numerically how the washout is kept under control in this family of
    models with more than one pair of RH neutrinos. The washout
    parameter $K_i$ is defined as
    \begin{equation}
    K_i= {\Gamma_i\over H(z=1)}= {(hh^{\dagger})_{ii} v_u^2 \over
    m_{\star}M_i} \,,
    \end{equation}
    where $H(z)$ is the usual Hubble expansion rate: $H(z)\simeq 1.66\sqrt{g_*}
    \frac{M_i^2}{z^2 M_{\rm Pl}}$, and $m_{\star}\simeq
    1.08\times 10^{-3}$~eV\footnote{Note here that for simplicity,
    we have assumed the SUSY breaking scale to be above the lightest RH neutrino
    mass so that only the SM degrees of freedom are in relativistic thermal
    equilibrium, i.e. $g_*\simeq 106.75$. However, the main results of this paper
    remain unchanged irrespective of the sparticle spectrum chosen.}. Plugging in
    numbers, we find that with Yukawa couplings of order $10^{-1}$ and a
    RH neutrino mass of order 1 TeV, the washout parameter $K$ is of order
    $10^{12}$, which is huge! However, the suppression of the washout is
    also very large, being proportional to $\delta^2$ with $\delta\ll 1$ due
    to the smallness of $\mu$, as required to get the right scale for the light
    neutrinos. Specifically, for the example of Eq.~(\ref{eq:murhs}),
    we find that $\delta\sim 10^{-5}$ and therefore the
    effective washout parameter $K^{\rm eff}\simeq \delta^2K \sim 100$, which is
    reasonably small.

    In the LR model we are considering, there are other processes
    contributing to the washout of lepton number, for instance, $N_R
    e_R\leftrightarrow \bar{u}_R d_R$. More precisely, this process
    destroys RH lepton number, but in the temperature range of interest
    to us (TeV scale) every individual RH lepton flavor equilibrates
    with the LH lepton flavor one, thanks to the Yukawa interactions.
    Does this process also turn off in the limit of lepton number
    conservation? It can be easily shown that, including the production
    of the RH neutrino by an inverse decay, followed by the scattering
    process mentioned above, there is also a destructive interference
    within the quasi-Dirac pair which leads exactly to the same kind of
    $\delta^2$-suppression as for the process $\ell \Phi \to
    \bar{\ell}\Phi^{\dagger}$.

    Another feature of inverse seesaw models is that they typically lead
    to lepton flavor equilibration~\cite{AristizabalSierra:2009mq}
    because of the large Yukawa couplings. More precisely, it can be
    shown that the process $\ell_{\alpha}\Phi\leftrightarrow
    \ell_{\beta}\Phi$, which does not change lepton number, but changes
    lepton flavor, is deep in thermal equilibrium for the TeV
    temperatures (see, for instance, Ref.~\cite{Antusch:2009gn}). Consequently,
    the Boltzmann equations for leptogenesis can be written as only one
    equation for the sum of the lepton
    flavors~\cite{AristizabalSierra:2009mq}. In other words, flavor
    effects~\cite{flavor} are not important in our framework.

    Putting together all the qualitatively important effects discussed
    above and solving the relevant set of Boltzmann equations, one can
    derive the following expression for the efficiency factor (see, for
    instance, Ref.~\cite{pedestrian}):
    \begin{eqnarray}
        \kappa_{i}(z) &\simeq& \int_{z_0}^{z} dz' \frac{dN^{\rm eq}_{N_i}(z')}{dz'}
        \frac{D(K_i,z')}{D(K_i,z')+D_{W_R}(z')+4S_{Z'}N_{N_i}^{\rm eq}({z'})+S_{W_R}
        (z')}\nonumber\\
        &&\times {\rm exp}\left[-\int_{z'}^{z}dz''\left\{\sum_i W_{\rm ID}(K_{i},z'')
        +W_{W_R}(z'')\right\}\delta_i^2\right] \, ,
        \label{eq:efficiency}
    \end{eqnarray}
    where $z\equiv M_i/T$,~ $N_X$ is the number density of $X$ over the
    relativistic number density of RH neutrinos, and $D,~S,~W$ denote
    the various decay, scattering and washout terms respectively,
    defined in Section IV. Note that the expression above assumes
    $dN_{N_i}/dz \simeq dN^{\rm eq}_{N_i}/dz$ with
    $   N_{N_i}^{\rm eq} = \frac{1}{2}z^2{\cal K}_2(z)$,
    ~where ${\cal K}_2(z)$ is the modified Bessel function of the 2nd kind.
    This is a very good approximation in our model (with large Yukawa couplings).
    Note also that we are neglecting spectator processes~\cite{spec} and
    $\Delta L=1$ scatterings
    involving the Higgs, which are both expected to lead to order one
    corrections.

    The final baryon asymmetry can be conveniently written as
    \begin{eqnarray}
        \eta_B \simeq 10^{-2}\sum_{i}\epsilon_{i}
    \kappa_{i}(z\to \infty)\, ,
    \label{eq:etab}
    \end{eqnarray}
    where the dilution factor $10^{-2}$ takes into account the fraction of $B-L$
    asymmetry converted into baryon asymmetry by sphaleron processes and also the
    dilution due to photon production from the onset of leptogenesis till
    recombination. $\epsilon_{i}$ is the $C\!P$-asymmetry generated by the decay
    of $N_i$ into any lepton flavor and is given by~\cite{Covi:1996wh}
    \begin{eqnarray}
        \epsilon_{i}=\frac{1}{8\pi}\sum_{j\neq i}
        { {\rm Im}\left[(hh^{\dagger})^2_{ij}\right]\over \sum_\beta |h_{i\beta}|^2}f^v_{ij}
        \label{eq:eps}
    \end{eqnarray}
    where $f^v$ is the $L$-violating self-energy and vertex loop
    factor\footnote{Note that the $L$-conserving self-energy contribution
    vanishes when one sums over flavor.}. In the quasi-degenerate limit
    of the $(i,j)$ pair, we have
    \begin{eqnarray}
        f^v_{ij}\simeq \frac{M_j^2-M_i^2}{(M_j^2-M_i^2)^2+(M_j\Gamma_j
        -M_i\Gamma_i)^2}\,.
    \end{eqnarray}

    Note that Eq.~(\ref{eq:eps}) was derived assuming heavy neutrino
    mass eigenstates. Therefore, it is necessary to make a basis
    transformation from the ``flavor'' basis where
    \begin{eqnarray}
        {\cal M}_{\rm RH} = \left(\begin{array}{cc}
        0 & M_N\\
        M_N^T & \mu
        \end{array}\right),
        \label{eq:rhmass}
    \end{eqnarray}
    to the diagonal mass basis with real and positive eigenvalues
    $M_i~(i=1,2,\ldots,6)$, grouped into three quasi-degenerate pairs
    with mass splittings in each pair of order $\mu_{kk}~(k=1,2,3)$. Analytically,
    the exact diagonalization of the full $6\times 6$ mass matrix ${\cal
    M}_{\rm RH}$ to get a closed form expression for the Yukawa
    couplings, $h_{i\alpha}$, in terms of the known parameters, namely
    $M_D,~M_N$ and $\mu$, is extremely involved. In Appendix B, we show the
    analytical expressions up to first order in $\mu$ for some simpler cases with
    only two quasi-Dirac pairs and show explicitly that the $C\!P$-asymmetry indeed
    vanishes in the $L$-conserving limit $\mu\to 0$, as expected.
    For the general case with three quasi-Dirac pairs, we numerically evaluate
    the $C\!P$-asymmetry in the next section. We note that
    the three-pair case reduces to the two-pair case discussed in
    Appendix B if one of the masses is much heavier than the other two and
    hence decouples from the rest.
    \section{Results}

    As noted in Section II, the Yukawa couplings are fixed by the
    $SO(10)$ symmetry. The $\mu$ matrix can be deduced from the knowledge of the
    light neutrino masses and mixing angles as a function of the RH neutrino mass
    matrix $M_N$, which can be taken to be diagonal without loss of generality.
     Varying the RH neutrino mass eigenvalues input then leads to
    different $\mu$ matrices, keeping the light neutrino mass matrix
    given by Eq.~(\ref{eq:lhmass}) such that its mass eigenvalues and
    mixing angles are within $2\sigma$ of the observed values. Once we
    know the explicit form of the RH neutrino mass matrix given by
    Eq.~(\ref{eq:rhmass}), we can define the quasi-Dirac pairs by
    transforming to a basis in which this mass matrix is diagonal with
    real and positive eigenvalues. We then calculate the
    $C\!P$-asymmetry and efficiency factors for the decay of the
    lightest RH neutrino pair and scan the parameter space to match the
    calculated baryon asymmetry (using Eq.~(\ref{eq:etab})) with the
    observed 68\% C.L. value, $\eta_B=(6.2\pm 0.15)\times 10^{-10}$~\cite{wmap}.
    Note that we only consider the asymmetry generated by decay of the
    lightest RH neutrino pair as the asymmetry generated by the heavy pairs
    is washed out very rapidly (due to large exponential suppression), and for
    these washouts not to affect
    the asymmetry generated by the lightest pair, we require the lightest
    pair to be at least 3 times smaller than the next heavy pair~\cite{Blanchet:2006dq}.

    To calculate the efficiency factor given by Eq.~(\ref{eq:efficiency}), we
    first write down the thermally averaged rates for $N\to \ell \Phi$ decay and
    the corresponding
    inverse decay~\cite{pedestrian}:
    \begin{eqnarray}
        D(K_i,z)&=&\frac{{\cal K}_1(z)}{{\cal K}_2(z)}K_iz,\nonumber\\
        W_{\rm ID}(K_i,z) &=&\frac{1}{4}K_i{\cal K}_1(z)z^3.
    \end{eqnarray}
    with ${\cal K}_a,~a=1,2$ denoting the modified Bessel
    function of the $i$th type.
    The thermally averaged rate $D_{W_R}$ for the $W_R$-mediated
    $N$-decay, $N\to \ell q_R\bar{q}'_R$, is given by
    \begin{eqnarray}
        D_{W_R}(z) = \frac{\gamma_N^{(W_R)}}{n^{\rm eq}_NHz}
        \label{eq:thermavg}
    \end{eqnarray}
    where $n_N^{eq}$ is the RH neutrino equilibrium number density,
    $n_N^{\rm eq}(z) = \frac{3}{4}n_\gamma(z)N_{N_i}^{\rm eq}(z)$ with
    $n_\gamma = \frac{2\zeta(3)}{\pi^2}T^3$, and $\gamma_N$ is the
    reaction density:
    \begin{eqnarray}
        \gamma_N^{(W_R)} = n_N^{\rm eq}(z)\frac{{\cal K}_1(z)}{{\cal K}_2(z)}
        \Gamma_N^{(W_R)},
    \end{eqnarray}
    where $\Gamma_N^{(W_R)}$ is the total three body decay width of $N$,
    given by~\cite{Frere:2008ct}
    \begin{eqnarray}
        \Gamma_N^{W_R} =
        \frac{3g_R^4}{2^9\pi^3M_N^3}\int_0^{M_N^2} ds
        \frac{(M_N^6-3M_N^2s^2+2s^3)}{(s-M_{W_R}^2)^2+M_{W_R}^2\Gamma_{W_R}^2}\,.
    \end{eqnarray}
    with the $W_R$ total decay width $\Gamma_{W_R} = \frac{g_R^2}{4\pi} M_{W_R}$.

    The various scattering rates $S_{W_R,Z'}$ appearing in
    Eq.~(\ref{eq:efficiency}) are also defined as in
    Eq.~(\ref{eq:thermavg}) where the corresponding scattering reaction
    density is related to the reduced cross section as follows (see, for
    instance, Ref.~\cite{Giudice:2003jh}):
    \begin{eqnarray}
        \gamma(ab\leftrightarrow cd) = \frac{M_N^4}{64\pi^4 z}\int_{x_{\rm thr}}^{\infty}
        dx \sqrt{x} \hat{\sigma}(x){\cal K}_1(z\sqrt{x})
    \end{eqnarray}
    with $x=s/M_N^2$ and the threshold value $x_{\rm
    thr}=\frac{1}{M_N^2} {\rm max}[(m_a+m_b)^2,(m_c+m_d)^2]$. The
    reduced cross sections for various $W_R$ exchange diagrams were
    computed in Ref.~\cite{Frere:2008ct}:
    \begin{eqnarray}
        \hat{\sigma}_{Ne_R\leftrightarrow \bar{u}_Rd_R}(x) &=& \frac{9g_R^4}{48\pi x}
        \frac{(1-3x^2+2x^3)}{\left[\left(x-\frac{M_{W_R}^2}{M_N^2}\right)^2+
        \frac{M_{W_R}^2\Gamma_{W_R}^2}{M_N^4}\right]}\, ,\\
        \hat{\sigma}_{N\bar{u}_R\leftrightarrow e_R\bar{d}_R} (x) &=& \frac
        {9g_R^4}{8\pi x} \int_{1-x}^0 du\frac{(x+u)(x+u-1)}
        {\left(u-\frac{M_{W_R}^2}{M_N^2}\right)^2}\, ,\\
        \hat{\sigma}_{Nd_R\leftrightarrow e_Ru_R}(x) &=&
        \frac{9g_R^4}{8\pi}\frac{M_N^2}{M_{W_R}^2}\frac{(1-x)^2}{\left(
        x+\frac{M_{W_R}^2}{M_N^2}-1\right)}\, ,
    \end{eqnarray}
    Here we have ignored the $t$-channel process $NN\to \ell \ell$ as the rate for this process falls off very rapidly for the region of interest, viz.
    $z>1$~\cite{Frere:2008ct}.

    The reduced cross section for the $Z'$ exchange diagram is given
    by~\cite{Plumacher:1996kc}
    \begin{eqnarray}
        \hat{\sigma}_{NN\leftrightarrow \ell\bar{\ell},q\bar{q}}(x) = \frac{13g_{B-L}^4}{6\pi}\frac{\sqrt{x(x-4)^3}}
        {\left(x-\frac{M_{Z'}^2}{M_N^2}\right)^2+
        \frac{M_{Z'}^2\Gamma_{Z'}^2}{M_N^4}}\,,
    \end{eqnarray}
    with the total $Z'$ decay width
    \begin{eqnarray}
        \Gamma_{Z'} = \frac{g_{B-L}^2}{24\pi}M_{Z'}\left[13+3\left(1-\frac
        {4M_N^2}{M_{Z'}^2}\right)^{3/2}\right]\,.
    \end{eqnarray}

    Before calculating the efficiency factor, it is instructive to
    compare all the reaction rates appearing in
    Eq.~(\ref{eq:efficiency}) to get a clear idea of various
    contributions. As an illustration, we consider the case with the RH
    Majorana neutrino mass eigenvalues $(3.5,3,1)$ TeV
    (as in Eq.~(\ref{eq:murhs})). The
    flavor-summed washout parameter for the decay of the lightest
    quasi-Dirac pair in this case is given by $K_3\simeq 4\times
    10^{11}$ whereas the effective washout parameter is given by
    $K_3^{\rm eff}\simeq \delta_3^2K_3\simeq 168$ which is reasonable.
    For comparison, the corresponding values for the two heavy pairs are
    $K_1^{\rm eff}\simeq K_2^{\rm eff} = 8\times 10^7$ which, when
    exponentiated in the washout term in Eq.~(\ref{eq:efficiency}),
    leads to a huge suppression, thus making the efficiency in those
    channels practically negligible. Hence, from now on, we will
    consider the decay of only the lightest pair.

    \begin{figure}[h!]
        \centering
        \includegraphics[width=10cm]{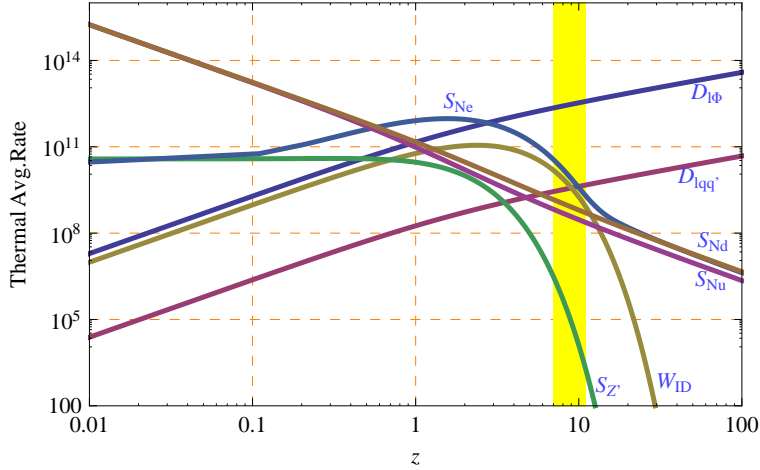}
        \caption{Various $N$ decay and scattering rates (thermally averaged)
        as a function of $z=M_{N_3}/T$ for a particular choice of RH neutrino
        masses, $(M_{N_1},M_{N_2},M_{N_3})=(3.5,3,1)$ TeV and
        $M_{W_R}=M_{Z'}=2$ TeV. The yellow shaded region is where the
        asymmetry is generated.}
        \label{fig:rate}
    \end{figure}
    In Fig.~\ref{fig:rate}, we show the various thermally averaged decay
    and scattering rates as a function of $z\equiv M_{N_3}/T$, for the above
    choice of the RH neutrino masses and for $M_{W_R}=M_{Z'}=2$ TeV. The yellow
    shaded region shows the asymmetry production time, approximately
    when $z_B-2< z < z_B+2$ with $z_B \simeq 2+4K^{0.13}e^{-2.5/K}
    \simeq 9$~\cite{Antusch:2009gn}\footnote{Here we have assumed that the
    production of asymmetry stops immediately after the temperature drops below the
    sphaleron freeze-out temperature, $T_{\rm sph}\simeq 130$~GeV for a
    Higgs mass $m_H=120$~GeV~\cite{Burnier:2005hp}.}. We note that in
    this range, the $N\to \ell \Phi$ decay rate, $D_{\ell\Phi}$,
    dominates over the three-body decay rate as well as all the
    scattering rates by several orders of magnitude. Hence in the
    efficiency factor, Eq.~(\ref{eq:efficiency}), the dilution term
    $D/(D+S)$ is very close to unity and is essentially independent of
    $M_{W_R}$ and $M_{Z'}$. The enhanced $N\to \ell\Phi$ decay rate is
    due to the large Yukawa couplings in the inverse seesaw scenario. We
    also note that as the $W_R$-mediated three-body $N$ decay rate is
    much smaller than the $N\to \ell \Phi$ decay rate, the washout term
    $W_{W_R}$ in Eq.~(\ref{eq:efficiency}) arising due to the process
    $\ell \Phi\to N\to \bar{\ell} q\bar{q}'$ which is proportional to
    the branching ratio of $N\to \bar{\ell} q\bar{q}'$ will be suppressed
    compared to the inverse decay term $W_{\rm ID}$. Thus we find that
    the efficiency factor is also essentially independent of both $W_R$ and
    $Z'$ masses for a wide range of parameter space. Of course, the
    $W_R$ and $Z'$ scattering terms will start to dominate for very low
    values of their masses; however, we estimated this lower bound to be
    well below the current collider bounds on $M_{W_R}$ and $M_{Z'}$ which are
    roughly a TeV or so~\cite{WRbound}.

 \begin{figure}[h!]
        \centering
        \includegraphics[width=10cm]{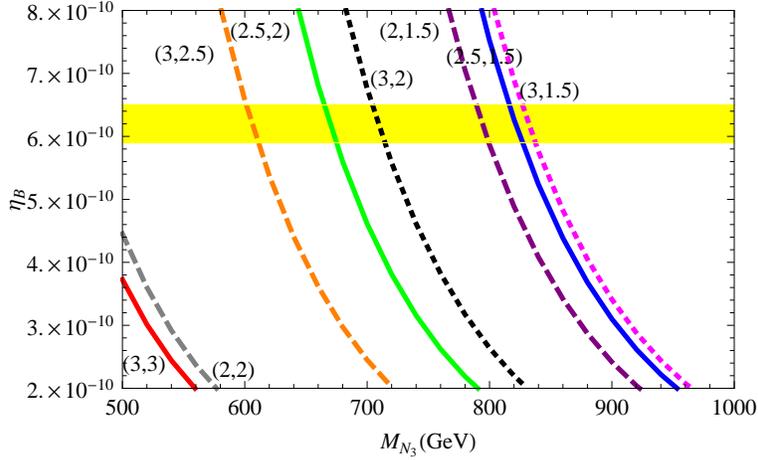}
        \caption{Correlation between the baryon asymmetry and the
        lightest RH neutrino mass for various heavy mass pairs $(M_{N_1},
	M_{N_2})$ in TeV. 
        The yellow shaded region is the observed value of $\eta_B$ within
        $2\sigma$ C.L..}
        \label{fig:eta}
    \end{figure}

    Fig.~\ref{fig:eta} shows the baryon asymmetry $\eta_B$ as a function of the
    lightest RH neutrino mass $M_{N_3}$ for different choices of the heavy
    RH neutrino masses $(M_{N_1},M_{N_2})$ from $1.5-3$ TeV. The
    calculated value of $\eta_B$ is to be compared with the observed 68\%
    C.L. value, $\eta_B=(6.2\pm 0.15)\times 10^{-10}$~\cite{wmap}. It is clear
    from the figure that for a given heavy mass
    pair, there is a narrow range of values allowed for $M_{N_3}$ satisfying 
    the  observed baryon asymmetry (the yellow shaded region). We note that for fixed $M_{N_1}$, the
    allowed range of $M_{N_3}$ decreases with increasing $M_{N_2}$, while for fixed $M_{N_2}$,
    the allowed range of $M_{N_3}$ increases with increasing $M_{N_1}$.
Also note that
    when the heavy pairs have degenerate mass, the baryon asymmetry gets
    suppressed (e.g. the lower two lines in Fig.~\ref{fig:eta})
    due to the suppression in the $C\!P$-asymmetry. Finally, we note
    that for a given set of heavy mass pairs, $\eta_B\propto M_{N_3}^{-3}$.

\begin{figure}[h!]
        \centering
        \includegraphics[width=10cm]{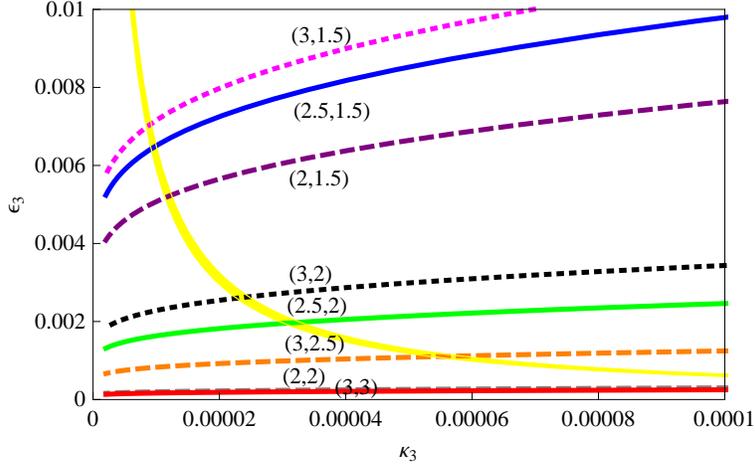}
        \caption{Correlation between the efficiency factor and the flavor-summed
        $C\!P$-asymmetry for the lightest pair for various values of the heavy mass
	pair $(M_{N_1},M_{N_2})$ in TeV. 
        The yellow shaded region corresponds
        to the observed value of $\eta_B$ within $2\sigma$ C.L..}
        \label{fig:ke}
    \end{figure}

    Fig.~\ref{fig:ke}  shows the correlation between the efficiency
    factor and the flavor-summed $C\!P$ asymmetry for various channels. The
    different lines correspond to different values of the heavy mass pair,
    $(M_{N_1},M_{N_2})$ in TeV, starting from (3,1.5) TeV at the top to (3,3) TeV at
    bottom. We note that for fixed $M_{N_1}$, the lines move
	down as we increase $M_{N_2}$ while for fixed $M_{N_2}$, they move
	up with increasing $M_{N_1}$.
The yellow shaded region shows the observed value of $\eta_B$
    which is
    essentially the product of $\kappa$ and $\epsilon$, summed over all
    pairs. As we have pointed out earlier, only the lightest pair
    contribution is significant, while the efficiency is too small for
    the other two pairs.

\section{Conclusion}
 In summary, we have shown that a TeV-scale Left-Right symmetry can be compatible
 with the understanding of the origin of matter via leptogenesis provided small neutrino
 masses are understood using the inverse seesaw mechanism. A crucial feature of this mechanism is that
 the magnitude of the lepton-number-breaking Majorana mass term is directly
 proportional to the neutrino mass, rather than inversely as in the usual type-I seesaw
 framework. This allows the Yukawa couplings that generate the Dirac mass for the neutrinos close
 to one, even with TeV-scale RH neutrinos. These two facts help to keep the wash-out of
 the generated lepton asymmetry under control, and thus explain the origin of matter while keeping both
 the $Z'$ and $W_R$ in the TeV range. The results of this paper should provide
 new motivation for searching for the $W_R$ and the Left-Right $Z'$ at the
 LHC. As has been already emphasized in literature~\cite{LHCWR}, the signal
 for the inverse seesaw with $W_R$ would be presence of trilepton final states with missing energy.
\section*{Acknowledgments}
The work of R.N.M. is supported by the National Science Foundation Grant No.
PHY-0968854. This work of S.B. has been partially supported by MICNN, Spain,
under contracts FPA 2007-60252 and Consolider-Ingenio CPAN CSD2007-00042 and
by the Comunidad de Madrid through Proyecto HEPHACOS ESP-1473. S.B.
acknowledges support from the CSIC grant JAE-DOC.

\appendix
\section{A New Coupling Unification Scenario}
\begin{figure}[h!]
    \centering
    \includegraphics[width=10cm]{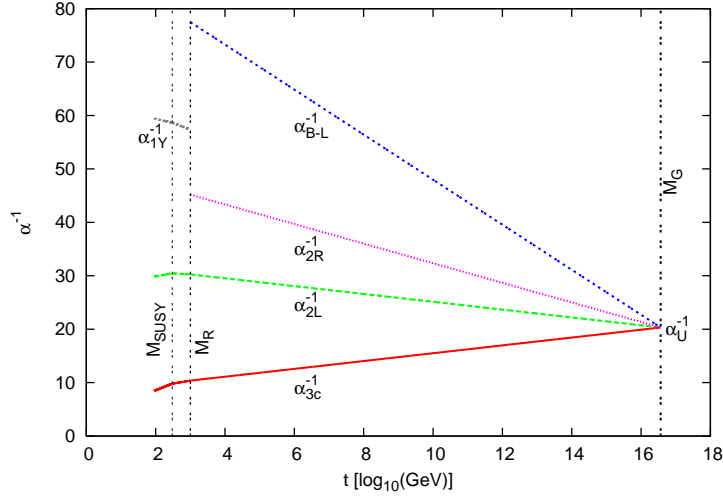}
    \caption{Gauge coupling unification in the new $SO(10)$ scenario with
    one $SU(2)_R$ doublet and one $SU(2)_R$ triplet.}
    \label{fig:uni}
\end{figure}

It was shown in Ref.~\cite{dev} that the $SO(10)$ embedding of
TeV-scale inverse seesaw mechanism discussed in this paper is
consistent with gauge coupling unification. In particular, it was
shown that for the symmetry breaking chain
\begin{eqnarray}
    SO(10)\stackrel {M_G} \longrightarrow {\bf 3}_c{\bf 2}_L{\bf 2}_R{\bf 1}
    _{B-L}~({\rm SUSYLR)}~\stackrel {M_R} \longrightarrow {\bf 3}_c{\bf 2}_L{\bf 1}_Y~({\rm MSSM})
    ~\stackrel {M_S} \longrightarrow {\bf 3}_c{\bf 2}_L
    {\bf 1}_Y~({\rm SM})~\stackrel {M_Z} \longrightarrow {\bf 3}_c{\bf 1}_Q,
    \nonumber
\end{eqnarray}
and for TeV-scale SUSY and $B-L$ breaking, it is possible to obtain gauge
coupling unification with two ${\bf 10}_H$ bidoublets, two RH ${\bf 16}_H$
doublets and one ${\bf 45}_H$ color triplet. In this section, we show that
there is an alternative choice of Higgs fields which also leads to coupling
 unification with TeV-scale $W_R$ and $Z'$. If we
add one set of $SU(2)_R$ triplets, $\Delta(1,1,3,0)$, coming from
the ${\bf 45}_H$ field, then we can also achieve unification with
the same field content as above except that we need only one
$SU(2)_R$ doublet. The SUSYLR $\beta$-function in this model is
given by
\begin{eqnarray}
    b_i = \left(10+\frac{3}{2}n_L+\frac{3}{2}n_R,n_{10}+n_L,n_{10}+n_R+2,-2
    \right)=\left(\frac{23}{2},2,3,-2\right)~ ~ (i={\bf 1}_{B-L},{\bf 2}_L,
    {\bf 2}_R,{\bf 3}_c)\nonumber
\end{eqnarray}
for $n_{10}=2,~n_L=0$ and $n_R=1$. As shown in Fig.~\ref{fig:uni},
we achieve unification at $M_G\simeq 4\times
10^{16}$ GeV with $\alpha_U^{-1}\simeq 20.3$.

This model has two nice features over the previous one: (i) all the Higgs
fields required for unification are connected to breaking of separate
gauge symmetries and there is no arbitrariness in
the number of fields, and (ii) the presence of the $SU(2)_R$ triplet enables
us to decouple the mass scales $M_{W_R}$ and $M_{Z'}$ which are otherwise
related in usual Left-Right models with $M_{Z'}>M_{W_R}$.
\section{Analytical Expression for the $C\!P$-Asymmetry}
The full Majorana mass matrix in the flavor basis $(N_i,S_i)$ is given by
\begin{eqnarray}
    {\cal M} = \left(\begin{array}{cc}
        0 & M_N\\
        M^T_N & \mu
    \end{array}\right)
\end{eqnarray}
where for $i=1,2,3$, both $M_N$ and $\mu$ are $3\times 3$ symmetric matrices. The Yukawa
Lagrangian in this basis is given by (with $i,j=1,2,3$)
\begin{eqnarray}
    {\cal L}_{y} = y_{i\alpha}\overline{N}_i\Phi^\dagger l_\alpha
    +M_{N_{ij}}N_i^T C^{-1}S_j+\frac{1}{2}\mu_{ij}S_i^T C^{-1}S_j+~{\rm h.c.}
\end{eqnarray}

In order to calculate the $C\!P$ asymmetry in this framework, it is
more convenient to work in the basis in which the RH Majorana
neutrino mass matrix is diagonal with real and positive eigenvalues.
The Lagrangian in this basis is given by (with $i=1,2,\cdots,6$)
\begin{eqnarray}
    {\cal L}_{h} = h_{i\alpha}\overline{\tilde{N}}_i\Phi l_{\alpha}+
    \frac{1}{2}M_{i}\tilde{N}_i^TC^{-1}\tilde{N}_i+{\rm h.c.}
\end{eqnarray}
Analytically, the exact diagonalization of the full $6\times 6$ mass
matrix ${\cal M}$ is extremely involved and we cannot obtain a
closed form expression for the $C\!P$-asymmetry in this case.
However, we can study the dependence of the small $L$-violating
parameter $\mu$ in some special cases, viz. when the $\mu$-matrix is
completely diagonal or completely off-diagonal, as in these cases
the Majorana mass matrix reduces to a block diagonal form. In this
section, we derive the analytical expression for the
$C\!P$-asymmetry in these two limits and for two sets of RH
neutrinos, i.e. for $(N_i,S_i)$ with $i=1,2$. The $i=3$ case reduces
to this limit if one of the masses is much heavier and hence
decouples from the other two.

We consider the $4\times 4$ version of the Majorana mass matrix ${\cal M}$:
\begin{eqnarray}
    {\cal M}_{4\times 4} = \left(\begin{array}{cc}
        0 & M_{N_{2\times 2}}\\
        M_{N_{2\times 2}} & \mu_{2\times 2}
    \end{array}\right)\, ,
\end{eqnarray}
where without loss of generality we choose the mass matrix
$M_N$ to be diagonal with real positive eigenvalues $M_{N_{1,2}}$.
However, the
elements of the $\mu$-matrix are, in general, complex quantities.
Now we consider two special cases:
\subsection*{\underline{\bf Case-I: $\mu$ purely diagonal}}

In this case, the Majorana mass matrix can be reduced to a simple
block diagonal form which decouples the $(N_1,S_1)$ and $(N_2,S_2)$
sectors:
\begin{eqnarray}
    {\cal M} = \left(\begin{array}{cccc}
        0 & 0 & M_{N_1} & 0\\
        0 & 0 & 0 & M_{N_2}\\
        M_{N_1} & 0 & \mu_{11} & 0\\
        0 & M_{N_2} & 0 & \mu_{22}
    \end{array}\right) \stackrel{r_2\leftrightarrow r_3,c_2\leftrightarrow c_3}\longrightarrow
\left(\begin{array}{cccc}
        0 & M_{N_1} & 0 & 0\\
        M_{N_1} & \mu_{11} & 0 & 0\\
        0 & 0 & 0 & M_{N_2}\\
        0 & 0 & M_{N_2} & \mu_{22}
    \end{array}\right)\,.
\end{eqnarray}
Then in the $(N_i,S_i)$ flavor basis, we have the $2\times 2$ matrices
\begin{eqnarray}
    \tilde{M}_i = \left(\begin{array}{cc}
        0 & M_{N_i}\\ M_{N_i} & \mu_{ii}
    \end{array}\right) = \left(\begin{array}{cc}
        0 & M_{N_i}\\ M_{N_i} & \varepsilon_i M_{N_i} e^{{\rm i}\theta_i}
    \end{array}\right)\,,
\end{eqnarray}
where $\varepsilon_i\equiv \mu_{ii}/M_{N_i}\ll 1$. The $\tilde{M}_i$ is
diagonalized with real and positive eigenvalues by a unitary
transformation $U_i^T \tilde{M}_i U_i$ where
\begin{eqnarray}
    U_i = \left(\begin{array}{cc}
        -{\rm i}\cos{\alpha_i}e^{{\rm i}\theta_i/2} & \sin{\alpha_i}e^{{\rm i}\theta_i/2}
        \\
        {\rm i}\sin{\alpha_i}e^{-{\rm i}\theta_i/2} & \cos{\alpha_i}e^{-{\rm i}\theta_i/2}
    \end{array}\right)\,,
    \label{eq:ui}
\end{eqnarray}
and the mixing angles are given by
\begin{eqnarray}
    \cos{\alpha_i} \simeq \frac{1}{\sqrt{2}}\left(1+\frac{\varepsilon_i}{4}\right),~
    ~ \sin{\alpha_i}\simeq
    \frac{1}{\sqrt{2}}\left(1-\frac{\varepsilon_i}{4}\right)\,,
\end{eqnarray}
up to ${\cal O}(\varepsilon_i)$. The corresponding mass eigenvalues are
given by
\begin{eqnarray}
    M_j \simeq M_{N_i}\left(1\pm \frac{\varepsilon_i}{2}\right)~ ~ (i=1,2;~
    j=1,2,3,4)\,.
\end{eqnarray}
It is clear that the mass splitting within a quasi-Dirac pair is given by
$\mu_{ii}$.

The Yukawa couplings in this diagonal mass basis are related to the
couplings in the flavor basis as follows:
\begin{eqnarray}
    h_{1\alpha} &\simeq&     \frac{{\rm i}e^{-{\rm i}\theta_1/2}}{\sqrt{2}}\left(1+\frac{\varepsilon_1}{4}    \right)y_{1\alpha},\nonumber\\
    h_{2\alpha} &\simeq &
    \frac{e^{-{\rm i}\theta_1/2}}{\sqrt{2}}\left(1-\frac{\varepsilon_1}{4}\right)
    y_{1\alpha},\nonumber\\
    h_{3\alpha} &\simeq&     \frac{{\rm i}e^{-{\rm i}\theta_2/2}}{\sqrt{2}}\left(1+\frac{\varepsilon_2}{4}    \right)y_{2\alpha},\nonumber\\
    h_{4\alpha} &\simeq &
    \frac{e^{-{\rm i}\theta_2/2}}{\sqrt{2}}\left(1-\frac{\varepsilon_2}{4}\right)
    y_{2\alpha}.
\end{eqnarray}
Note that in the $L$-conserving limit $\varepsilon_i\to 0$, we have
$h_{i\alpha} = {\rm i} h_{j\alpha}$ within a quasi-degenerate pair $(i,j)$,
as expected.

Now let us calculate the $C\!P$-asymmetry for the decay of one of
the quasi-Dirac particles, say $i=1$. We have from
Eq.~(\ref{eq:eps}),
\begin{eqnarray}
    \epsilon_1=\frac{1}{8\pi}\sum_{j\neq 1}
    \frac{ {\rm Im}\left[(hh^\dagger)^2_{1j}\right]}
    {\sum_\beta |h_{1\beta}|^2}f^v_{1j}\simeq
    \frac{\varepsilon_2}{16\pi
    \sum_\beta |y_{1\beta}|^2}{\rm Im}\left[e^{ {\rm i}(\theta_1-\theta_2)}
    \left(\sum_\alpha y^*_{1\alpha}y_{2\alpha}\right)^2\right]f^v_{13}
    \label{eq:eps1}
\end{eqnarray}
assuming $f_{13}\simeq f_{14}$. Note that the $j=2$ term vanishes as there
is no imaginary part in that case. It is clear that $\epsilon_1$ vanishes
as $\mu_{22}\to 0$. Similarly, one can show that $\epsilon_2$ also vanishes
in the limit $\mu_{22}\to 0$, and $\epsilon_3,~\epsilon_4$ vanish as
$\mu_{11}\to 0$.
\subsection*{\underline{\bf Case II: $\mu$ purely off-diagonal}}

In this case, the Majorana mass matrix in the $(N_i,S_i)$ flavor basis reduces
to the following block diagonal form:
\begin{eqnarray}
    {\cal M} = \left(\begin{array}{cccc}
        0 & 0 & M_{N_1} & 0\\
        0 & 0 & 0 & M_{N_2}\\
        M_{N_1} & 0 & 0 & \mu_{12}\\
        0 & M_{N_2} & \mu_{12} & 0
    \end{array}\right) \stackrel{c_1\leftrightarrow c_3,r_2\leftrightarrow r_4}\longrightarrow \left(\begin{array}{cccc}
        M_{N_1} & 0 & 0 & 0\\
        \mu_{12} & M_{N_2} & 0 & 0\\
        0 & 0 & M_{N_1} & \mu_{12}\\
        0 & 0 & 0 & M_{N_2}
    \end{array}\right)\,,
\end{eqnarray}
which, however, mixes the (1,2) sectors; in the $(N_1,S_2)$ basis, we have the $2\times 2$ mass matrix
\begin{eqnarray}
    \tilde{M} = \left(\begin{array}{cc}
        M_{N_1} & 0\\
        \mu e^{{\rm i}\theta} & M_{N_2}
    \end{array}\right)\,,
\end{eqnarray}
with $\mu\ll M_{N_1},M_{N_2}$. However, unlike in Case I, we cannot diagonalize
this asymmetric matrix by a single unitary transformation; instead, we have to
apply a bi-unitary transformation of the form $V^\dagger \tilde{M} U$. We
find that the following forms of $U$ and $V$ diagonalize $\tilde{M}$:
\begin{eqnarray}
    U = \left(\begin{array}{cc}
        \cos{\alpha} & \sin{\alpha}\\
        -\sin{\alpha}e^{{\rm i}\theta} & \cos{\alpha}e^{{\rm i}\theta}
    \end{array}\right),~ ~ V=\left(\begin{array}{cc}
        \cos{\beta} & \sin{\beta}\\
        -\sin{\beta}e^{{\rm i}\theta} & \cos{\beta}e^{{\rm i}\theta}
    \end{array}\right)\,,
\end{eqnarray}
where the mixing angles are given by
\begin{eqnarray}
    \cos{\alpha} &=& \frac{M_{N_2}^2-M_{N_1}^2}
    {\sqrt{(M_{N_2}^2-M_{N_1}^2)^2+\mu^2 M_{N_2}^2}},\nonumber\\
    \sin{\alpha} &=& \frac{\mu M_{N_2}}{\sqrt{(M_{N_2}^2-M_{N_1}^2)^2+\mu^2 M_{N_2}^2}},\nonumber\\
    \cos{\beta} &=& \frac{M_{N_2}^2-M_{N_1}^2}{\sqrt{(M_{N_2}^2-M_{N_1}^2)^2
    +\mu^2 M_{N_1}^2}},\nonumber\\
    \sin{\beta} &=& \frac{\mu M_{N_1}}{\sqrt{(M_{N_2}^2-M_{N_1}^2)^2+\mu^2 M_{N_1}^2}}.
\end{eqnarray}

The eigenvalues of $\tilde{M}$ are given by
\begin{eqnarray}
    M_i \simeq \frac{1}{\sqrt{2}}\left[M_{N_1}^2+M_{N_2}^2+\mu^2\mp \sqrt{
    (M_{N_2}^2-M_{N_1}^2)^2+2\mu^2(M_{N_1}^2+M_{N_2}^2)}\right]^{1/2}\,,
\end{eqnarray}
up to order ${\cal O}(\mu^2)$. Note however that in the new basis,
the mass matrix is still not diagonal and is of the form
$\left(\begin{array}{cc} 0 & M_i\\ M_i& 0\end{array}\right)$
    with eigenvalues $\pm M_i$. This can be diagonalized with real and
    positive eigenvalues $|M_i|$ by another unitary transformation:
    \begin{eqnarray}
        U_d = \left(\begin{array}{cc}
            \cos{\frac{\pi}{4}} &
            \sin{\frac{\pi}{4}} \\
            -\sin{\frac{\pi}{4}} &
            \cos{\frac{\pi}{4}}
        \end{array}\right)\cdot {\rm diag}\left(e^{-{\rm i}\pi/2},1\right)
    \end{eqnarray}
We note here that in this case, unlike in case I, there is no mass splitting
within the pair and the two quasi-Dirac RH neutrinos are exactly degenerate.
This is a general result that the off-diagonal elements of $\mu$ do not
contribute to the mass splitting within a pair; they just shift the
eigenvalues. Hence, the splitting can be approximated
by the diagonal elements of $\mu$, as in Eq.~(\ref{eq:deltai}).

Finally, the Yukawa couplings in the mass-diagonal basis with real
and positive eigenvalues are given in terms
of the couplings in the flavor basis as follows:
\begin{eqnarray}
    h_{1\alpha} &=&
    \frac{\rm i}{\sqrt{2}}\left(\cos{\beta}~y_{1\alpha}+\sin{\alpha}e^{-{\rm i}\theta}
    ~y_{2\alpha}\right),\nonumber\\
    h_{2\alpha} &=&
    \frac{1}{\sqrt{2}}\left(\cos{\beta}~ y_{1\alpha}
    -\sin{\alpha}e^{-{\rm i}\theta}~ y_{2\alpha}\right),\nonumber\\
    h_{3\alpha} &=&
    \frac{\rm i}{\sqrt{2}}\left(\sin{\beta}~y_{1\alpha}-
    \cos{\alpha}e^{-{\rm i}\theta}~y_{2\alpha}\right),\nonumber\\
    h_{4\alpha} &=&
    \frac{1}{\sqrt{2}}\left(\sin{\beta}~y_{1\alpha}+
    \cos{\alpha}e^{-{\rm i}\theta}~y_{2\alpha}\right).
    \label{eq:nyukc2}
\end{eqnarray}
Note that in the $L$-conserving limit $\mu\to 0$,
$\cos{\alpha},\cos{\beta}\to 1$ and $\sin{\alpha},\sin{\beta}\to 0$; it
is clear from Eqs.~(\ref{eq:nyukc2}) that in this limit, we recover
the relation $h_{i\alpha} = {\rm i}h_{j\alpha}$ for the $(i,j)$ pair.

Using Eqs.~(\ref{eq:nyukc2}), it can be shown that the
$C\!P$-asymmetry, Eq.~(\ref{eq:eps}), for $i=1$ becomes
\begin{eqnarray}
    \epsilon_1 &=& \frac{1}{8\pi\sum_{\gamma}|h_{1\gamma}|^2}
    \left[\sin{\alpha}\cos{\beta}\left(\cos^2{\beta}
    \sum_\gamma|y_{1\gamma}|^2-\sin^2{\alpha}\sum_\gamma|y_{2\gamma}|^2
    \right)f^v_{12}\right.\nonumber\\
    && \left. -\frac{1}{2}\left\{\cos\alpha\sin\beta\left(\cos^2\beta
    \sum_\gamma|y_{1\gamma}|^2-
    \sin^2\alpha\sum_\gamma|y_{2\gamma}|^2\right)\right.\right.
    \nonumber\\
    && \left.\left.+
    \cos\alpha\cos\beta\left(\cos\beta\sin\beta\sum_\gamma|y_{1\gamma}|^2
    -\cos\alpha\sin\alpha\sum_\gamma|y_{2\gamma}|^2\right)\right\}f^v_{13}
    \right]{\rm Im}\left(e^{-{\rm i}\theta} \sum_\gamma
    y^*_{1\gamma}y_{2\gamma}\right)\nonumber\\
    \label{eq:eps2}
    \end{eqnarray}
which clearly vanishes in the limit $\mu\to 0$ (as $\sin\alpha,\sin\beta
\propto \mu$).
Similarly it can be shown for other channels.

Comparing the $C\!P$-asymmetries $\epsilon_1$ in these two cases,
we find that in
Case I, the contribution within the pair vanishes and the remaining term
in Eq.~(\ref{eq:eps1}) which is proportional to $f^v_{13}$ is highly
suppressed as $M_1$ is not
quasi-degenerate with the $(M_3,M_4)$ pair. On the other hand, in case II,
the dominant contribution comes from within the $(M_1,M_2)$ pair which is
enhanced due to large $f^v_{12}$. Hence, combining these results, we expect
that in the general case with both diagonal and off-diagonal $\mu$-entries,
the dominant contribution to the $C\!P$-asymmetry $\epsilon_i$ should come from
``within the pair'' decay of $N_i$. We checked numerically that this is indeed
the case.

\end{document}